\documentclass[aps,showpacs,twocolumn,pre]{revtex4}
\usepackage{amssymb}
\usepackage{natbib}
\usepackage{amsmath}
\usepackage{amsfonts}
\usepackage{graphicx}
\usepackage{mathrsfs}
\usepackage{dcolumn}
\usepackage{bm}
\usepackage{color}
\definecolor{rot}{rgb}{0.75,0.05,0.25}
\definecolor{hellgrau}{gray}{0.5}
\definecolor{blau}{rgb}{0,0,0.7}

\def\Tr{\mbox{Tr}}

\DeclareMathOperator{\artanh}{artanh}

\begin{document}

\title{Thermodynamics and Fluctuation Theorems for a Strongly Coupled  Open Quantum System: An Exactly Solvable Case}

\author{Michele Campisi}
\email{Michele.Campisi@physik.uni-augsburg.de}
\author{Peter Talkner}
\author{Peter H\"anggi}
\affiliation{Institute of Physics, University of Augsburg,
  Universit\"atsstrasse 1, D-86135 Augsburg, Germany}
\date{\today }

\begin{abstract}
We illustrate recent results concerning the validity of the work fluctuation theorem in open
quantum systems [M. Campisi, P. Talkner, and P. H\"{a}nggi, Phys. Rev. Lett.
{\bf 102}, 210401 (2009)], by applying them to a  solvable model of an open quantum system.
The central role played by the thermodynamic partition function of the open quantum system, --
a two level fluctuator  with a strong quantum nondemolition coupling to a harmonic oscillator --,
is elucidated. The corresponding quantum Hamiltonian of mean force is evaluated explicitly.
We study the thermodynamic entropy and the corresponding specific
heat of this open system as a function of temperature and coupling strength and show that
both may assume negative values at nonzero low temperatures.
\end{abstract}
 \maketitle
\section{Introduction}
Exact results about nonequilibrium fluctuations in nanosystems, such as the Jarzynski equality \cite{Jarz97}  and the Tasaki-Crooks fluctuation theorem \cite{Tasaki:2000pi,Crooks:1999fq} have recently attracted a great deal of attention in the burgeoning field of nonlinear fluctuation relations. These results were first derived for classical systems and later for quantum systems that are either isolated or weakly coupled to their environment \cite{Tasaki:2000pi,Mukamel:2003ip,De-Roeck:2004cq,Esposito:2006by,Andrieux08,CrooksJSM08,Talkner:2007sf,Talkner:2008mr,TH2007}. However, often the interaction with the environment does play an important role which cannot be neglected in real experimental situations.
For this reason more attention has been  recently devoted to the strong coupling regime, both classically \cite{Jarzynski:2004pv}, and quantum mechanically \cite{CHT-PRL-09}. In this regime, the driven system of interest (with Hamiltonian $\hat{H}_S(t)$), strongly couples to a bath ($\hat{H}_B$), via a non negligible interaction term $\hat{H}_{SB}$:
\begin{equation}
\hat{H}(t)=\hat{H}_S(t)+\hat{H}_{SB}+\hat{H}_{B} \; .
\label{eq:Ham}
\end{equation}
For the applicability of work and fluctuation theorems, the correct choice of the statistical mechanical description of an open quantum system in terms of the proper thermodynamic partition function, is of decisive importance. The bare system partition sum
\begin{equation}
Q_S(t)=\Tr_S{e^{-\beta \hat{H}_S(t)}}
\label{eq:Q(t)intr}
\end{equation}
clearly fails to account for the effects of the environment on the system. It rather is the open system thermodynamic partition function
\begin{equation}
Z_S(t) = \frac{Y(t)}{Z_B}
\label{eq:Zs=Z-Zb}
\end{equation}
which consistently accounts for these effects \cite{CHT-PRL-09,HIT_NJP08,Ingold:2008mz,Horhammer,Fundamentals,Haenggi:2006br,Ingold:2002pc,Theo2002,
HaenggiQTransp98,Grabert:1988et,Ford:1985it,Grabert:1984ux,Caldeira:1983aj,FeynStatMech,
Feynman:1963qm}. Here $Y(t)$ denotes the total system partition function; i.e.,
\begin{equation}
Y(t)={\Tr
  e^{-\beta(\hat{H}_S(t)+\hat{H}_{SB}+\hat{H}_{B})}}
\label{eq:Y(t)-intr}
\end{equation}
and $Z_B$ the bare bath partition function
\begin{equation}
Z_B={\Tr_{B} e^{-\beta
    \hat{H}_{B}}}
\label{eq:Z_B-intr}
\end{equation}
The time $ t $ merely specifies the values of the external parameters
as they occur in the course of the driving  protocol at the time $t$. The symbols $\Tr_S, \Tr_B, \Tr$ denote traces over system, bath, and total system respectively.  The symbol $\beta=(k_BT)^{-1}$ indicates the inverse thermal energy, with $k_B$ Boltzmann constant and $T$ the temperature. This temperature is provided via vanishingly small weak contact with a large (super)-bath, which allows for a statistical mechanical treatment. 

The adoption of the thermodynamic partition function $Z_S(t)$, and the corresponding free energy $F_S(t)=-\beta^{-1}\ln Z_S(t)$ allows to obtain the Jarzynski equality
\begin{equation}
 \langle e^{-\beta w} \rangle = e^{-\beta \Delta F_S}
 \label{eq:JE}
\end{equation}
valid irrespectively of the coupling strength \cite{CHT-PRL-09}.

In the following we exemplify this result by applying it to a simple model Hamiltonian of an open quantum system, Sec. \ref{sec:model}, \ref{sec:FT}. We next illustrate the equilibrium thermodynamics of that open system, by computing its Hamiltonian of mean force, its entropy and specific heat, see Sec. \ref{sec:EqTher}. Remarks and conclusions are drawn in Sec. \ref{sec:conclusions}.

\section{\label{sec:model} A Two Level Fluctuator-Oscillator Model}
We consider the following Hamiltonian describing a two level system and a harmonic oscillator interacting with each other:
\begin{equation}
\hat{H}(t)= \frac{\varepsilon(t)}{2} \hat{\sigma}_z \otimes \hat{1}_B + \hat{1}_S \otimes \Omega \left(\hat{a}^\dag  \hat{a}+\frac{1}{2}\right) + \chi  \hat{\sigma}_z \otimes \left(\hat{a}^\dag  \hat{a}+\frac{1}{2}\right).
\label{eq:H}
\end{equation}
Here $\hat{\sigma}_z$ is a Pauli matrix of the two level system, $\hat{a}^\dag$ and $\hat{a}$ are raising and lowering
operators of the harmonic oscillator, $\varepsilon(t),\Omega,\chi$ are the two level system energy spacing, the oscillator energy quantum and the coupling energy, respectively. The parameter $\chi$ can assume positive and negative values whereas $\varepsilon(t)$ and $\Omega$ are strictly positive. The oscillator energy quantum $\Omega$ is related to the oscillator frequency $\omega$, via Planck's constant $\Omega = \hbar \omega$. We consider the two level system (also referred to as the qubit throughout the text) as our system of interest ($\hat{H}_S(t)={\varepsilon(t)} \hat{\sigma}_z/2\otimes \hat{1}_B$), and the oscillator as our stylized, ``minimal'' bath; i.e. $\hat{H}_B=\Omega (\hat{a}^\dag \hat{a}+1/2)\otimes \hat{1}_S$.  The operators $\hat{1}_{S}$ and $\hat{1}_B$ denote the identity operators acting on the system and bath Hilbert spaces, respectively.
We require $|\chi|<\Omega$,
which ensures that the total Hamiltonian is bounded from belowguaranteeing stability of the total system (from Eq. (\ref{Ens}), $\Omega + \chi s$ must be positive in order that the smallest eigenvalue be finite).
The two level system energy spacing $\varepsilon$  is assumed to depend on time according to some pre-specified protocol. This model Hamiltonian has the peculiarity that  the interaction Hamiltonian commutes with both system and bath Hamiltonians, implying a so called quantum nondemolition coupling:
\begin{equation}
[\hat{H}_S(t) \otimes \hat{1}_B,\hat{H}_{SB}]=[\hat{1}_S\otimes \hat{H}_B,\hat{H}_{SB}]=0 \ .
\end{equation}

The time-instantaneous energy eigenvalues assume the form
\begin{equation}
E_{n,s}(t) = \frac{\varepsilon(t)}{2} s + \Omega \left(n+\frac{1}{2}\right) + \chi s \left(n+\frac{1}{2}\right)
\label{Ens}
\end{equation}
$s = \pm 1, \; n = 0,
1, 2, \ldots$ . 
We remark that the corresponding instantaneous eigenstates $|n,s\rangle$ do not depend on time.

The partition function $Y(t)$ of the total system becomes, with Eq. (\ref{eq:Y(t)-intr})
\begin{equation}
Y(t)= \sum_{s,n} e^{-\beta E_{n,s}(t)} = q_+(t)+q_-(t)
    \label{eq:Y(t)}
\end{equation}
where
\begin{equation}
q_{\pm}(t) = \frac{e^{-\beta \Omega/2} e^{\mp \beta (\varepsilon(t)+\chi)/2}}{1-e^{-\beta(\Omega \pm \chi)}} \ .
\end{equation}
The bare bath partition function $Z_B$ is, with Eq. (\ref{eq:Z_B-intr}):
\begin{equation}
Z_{B} = \sum_{n} e^{-\beta \Omega (n+1/2)}= \frac{1}{2\sinh( \beta \Omega/2)} \ .
\label{eq:Z_B}
\end{equation}
Then the thermodynamic partition function $Z_S(t)$ of the open system becomes, according to Eq. (\ref{eq:Zs=Z-Zb}):
\begin{equation}
Z_{S}(t) = 2(q_+(t)+q_-(t))\sinh(\beta \Omega/2) \ .
     \label{eq:Z_S(t)}
\end{equation}
Note that the open system thermodynamic partition function differs substantially from the bare system partition sum $Q_S(t)$, which reads, with Eq. (\ref{eq:Q(t)intr}):
\begin{equation}
Q_S(t) =  \sum_{s} e^{-\beta \varepsilon(t) s/2}
= 2 \cosh ( \beta {\varepsilon(t)}/{2}) \ .
     \label{eq:Q_S(t)}
\end{equation}
In particular the thermodynamic partition function  consistently accounts for the presence of the oscillator and the interaction, as it depends on $\Omega$ and $\chi$, whereas the partition sum $Q_S(t)$ does not.

\section{\label{sec:FT}Work and Fluctuation Theorems}
For a prescribed protocol of the two level spacing $\varepsilon(t)$ $t_0\leq t \leq t_f$, the work $w$ performed on the two level system is distributed according to the probability density function $p_{t_f, t_0}(w)$, given by \cite{Talkner:2007sf}:
\begin{equation}
\begin{split}
p_{t_f, t_0}(w)  =\sum_{m,n=0}^{\infty} \sum_{r,s=\pm1} & \delta(w-(E_{m,r}(t_f)-E_{n,s}(t_0))) \\ & \times  P(m,r|n,s)\frac{e^{-\beta E_{n,s}(t_0) }}{Y(t_0)}
\end{split}
\end{equation}
where $\delta(x)$ denotes the Dirac delta function and $P(m,r|n,s)$ is the transition probability to jump from the eigenstate $|n,s\rangle$ of the total Hamiltonian at time $t_0$ to the eigenstate $|m,r\rangle$ at time $t_f$:
\begin{equation}
P(m,r|n,s)= |\langle m,r| \hat{U}_{t_f,t_0} |n,s\rangle |^{2}
\label{eq:p(mr|ns)}
\end{equation}
with $\hat{U}_{t_f,t_0}=\mathcal{T}\exp (-i\int_{t_0}^{t_f} dt\hat{H}(t)/\hbar)$ denoting the time evolution operator. The model Hamiltonian in Eq. (\ref{eq:H}), commutes with itself at different times, so the time ordered exponential reduces to an ordinary exponential
\begin{equation}
\begin{split}
\hat{U}_{t_f,t_0} = \exp \left[-\frac{i}{\hbar} \left( \int_{t_0}^{t_f} dt \frac{\varepsilon(t)}{2} \hat{\sigma}_z \otimes \hat{1}_B  \right. \right. \\
+ \hat{1}_S \otimes \Omega  \left(\hat{a}^\dag  \hat{a}+\frac{1}{2}\right) (t_f-t_0)\\
+ \left. \left. \chi  \hat{\sigma}_z \otimes  \left(\hat{a}^\dag  \hat{a}+\frac{1}{2} \right)(t_f-t_0)  \right)\right]
\end{split}
\end{equation}
By inserting this expression into Eq. (\ref{eq:p(mr|ns)}), one sees that no transition takes place
\begin{equation}
P(m,r|n,s)= \delta_{m,n}\delta_{r,s}
\end{equation}
with $\delta_{m,n}$ denoting the Kronecker symbol. This is of course to be expected since an interaction that commutes with the free evolution does not cause any transition. Thus, for the work probability density one obtains:
\begin{equation}
\begin{split}
p_{t_f, t_0}&(w)  =  \\
& \frac{q_+(t_0)\delta (w-\Delta\varepsilon/2)+q_-(t_0)\delta(w+\Delta\varepsilon/2)}{q_+(t_0)+q_-(t_0)}
\end{split}
\label{eq:p(w)}
\end{equation}
where $\Delta\varepsilon=
\varepsilon(t_f)-\varepsilon(t_0)$.
By exchanging $t_0$ with $t_f$ one obtains the backward pdf of work $p_{t_0, t_f}(w)$, corresponding to the backward protocol $\bar{\varepsilon}(t)=\varepsilon(t_f+t_0-t)$. After some calculations one obtains the following expression for their ratios:
\begin{equation}
\frac{p_{t_f, t_0}(w)}{p_{t_0, t_f}(-w)} = e^{\beta w} \frac{\cosh(\beta \frac{\varepsilon(t_f)+\chi}{2})-e^{-\beta \Omega}\cosh(\beta \frac{\varepsilon(t_f)-\chi}{2})}{\cosh(\beta \frac{\varepsilon(t_0)+\chi}{2})-e^{-\beta \Omega}\cosh(\beta \frac{\varepsilon(t_0)-\chi}{2})}
\end{equation}
where we recognize that the ratio on the right hand side is equal to $Z_S(t_f)/Z_S(t_0)$, as predicted by the work fluctuation theorem for arbitrary open quantum systems \cite{CHT-PRL-09}.
Using (\ref{eq:p(w)}) we also obtain the following expression for the Jarzynski exponentiated work:
\begin{equation}
\langle e^{-\beta w}\rangle=\frac{\cosh(\beta \frac{\varepsilon(t_f)+\chi}{2})-e^{-\beta \Omega}\cosh(\beta \frac{\varepsilon(t_f)-\chi}{2})}{\cosh(\beta \frac{\varepsilon(t_0)+\chi}{2})-e^{-\beta \Omega}\cosh(\beta \frac{\varepsilon(t_0)-\chi}{2})}
\label{eq:JEb}
\end{equation}
that is,
\begin{equation}
\langle e^{-\beta w}\rangle=\frac{Z_S(t_f)}{Z_S(t_0)}
\label{eq:JE2}
\end{equation}
as predicted by the Jarzynski equality for arbitrary open quantum systems in Eq. (\ref{eq:JE}) \cite{CHT-PRL-09}.

By comparison of Eqs. (\ref{eq:Q_S(t)})  and (\ref{eq:JEb}) one
observes that:
\begin{equation}
 \langle e^{-\beta w} \rangle \neq \frac{Q_S(t_f)}{Q_S(t_0)}.
 \end{equation}
This result is in contrast to recent claims reported by Teifel and Mahler \cite{Teifel:2007qr},
according to which the averaged exponentiated work should be identical to the ratio of partition
sums $Q_S$ independently of coupling strength, provided the interaction commutes with both
system and bath Hamiltonians as it is the case with the present study.
\begin{figure}
\begin{center}
\includegraphics[width=8cm]{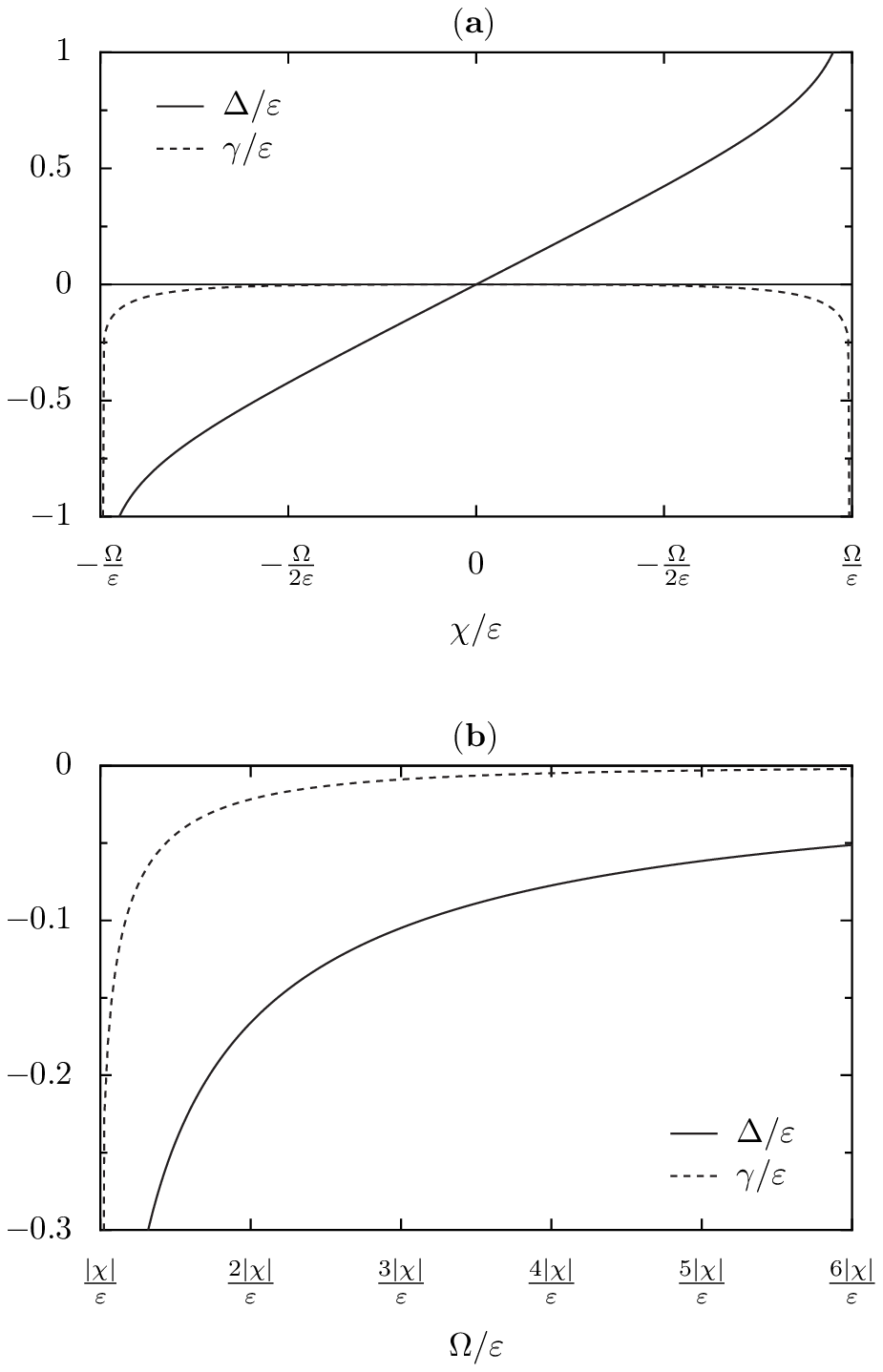}
 \caption{Dimensionless difference between renormalized qubit's energy spacing and original
 qubit's energy spacing, $\Delta/\varepsilon$ (solid line, Eqs. (\ref{eq:varepsilon*},\ref{eq:delta})), and global dimensionless shift
 of the energy spectrum, $\gamma/\varepsilon$ (dashed line, Eq. (\ref{eq:gamma})), as functions of (a) the
 dimensionless coupling strength $\chi/\varepsilon$ (top panel), (b) dimensionless oscillator's energy quantum $\Omega/\varepsilon$ (bottom panel).
The graphs correspond  to a temperature of $T=50$mK, and the experimental values  employed in Ref.  \cite{Schuster:2007ph} $\varepsilon/2\pi\hbar=6.9$ GHz,
 (a) $\Omega /2 \pi \hbar =5.7$GHz (top panel),  (b) $\chi /\pi \hbar =-17$MHz (bottom panel).}
\label{fig:1}
\end{center}
\end{figure}
\section{\label{sec:EqTher}Equilibrium Thermodynamics}
We turn now to the study of the equilibrium thermodynamics of the open two level system.
This means that we now keep $\varepsilon$ fixed and study the time independent Hamiltonian
\begin{equation}
\hat{H}= \frac{\varepsilon}{2} \hat{\sigma}_z \otimes \hat{1}_B +
\hat{1}_S \otimes \Omega \left(\hat{a}^\dag  \hat{a}+\frac{1}{2}\right) +
\chi  \hat{\sigma}_z \otimes \left(\hat{a}^\dag  \hat{a}+\frac{1}{2}\right).
\label{eq:H-time-ind}
\end{equation}

\subsection{The Hamiltonian of Mean Force}
A fundamental quantity that is closely related to the open system partition function $Z_S$ is the quantum Hamiltonian of mean force \cite{CHT-PRL-09}
\begin{equation}
\hat{H}^*:= -\frac{1}{\beta}\ln \frac{\Tr_B
  e^{-\beta(\hat{H}_S+\hat{H}_{SB}+\hat{H}_{B})}}{Z_B}.
    \label{eq:H*}
\end{equation}
It generalizes the potential of mean force commonly employed in reaction rate theory \cite{HTB1990} and implicit solvent models \cite{Roux19991}. The Hamiltonian of mean force is the effective Hamiltonian that describes the open system at equilibrium with the environment according to the equation:
\begin{equation}
Z^{-1}_{S} e^{-\beta \hat{H}^*}= Y^{-1} \Tr_B e^{-\beta
\hat{H}}\;.
\label{rhoS}
\end{equation}
It hence determines the reduced density matrix of the open system, $\rho_S$, in thermal equilibrium according to $\rho_S=Z^{-1}_{S} e^{-\beta \hat{H}^*}$.
The calculation of $\hat{H}^*$ in general is a difficult task. However for the model Hamiltonian in Eq. (\ref{eq:H-time-ind}), the calculation is straightforward and leads to
\begin{equation}
\hat{H}^* = \frac{\varepsilon^*}{2} \hat{\sigma}_z + \gamma  \hat{1}_S
\label{eq:H*special}
\end{equation}
where
\begin{equation}
\varepsilon^* = \varepsilon + \chi + \frac{2}{\beta} \artanh \left( \frac{e^{-\beta \Omega} \sinh(\beta \chi)}{1-e^{-\beta \Omega} \cosh(\beta \chi)} \right)
\label{eq:varepsilon*}
\end{equation}
is the renormalized level spacing, and
\begin{equation}
\gamma = \frac{1}{2 \beta}\ln \left( \frac{ {1-2e^{-\beta \Omega} \cosh(\beta \chi) + e^{-2 \beta \Omega}} }{(1-e^{-\beta \Omega})^2} \right)
\label{eq:gamma}
\end{equation}
specifies  a global shift of the spectrum.  In obtaining Eq. (\ref{eq:H*special}) we used the identity $e^{a \hat{\sigma}_z}=\cosh(a)\hat{1}_S+\sinh(a)\hat{\sigma}_z$.
When  $\chi\rightarrow 0$, the renormalized spacing tends to the original spacing $\varepsilon$, and the offset $\gamma$ vanishes, so that $H^*$ tends to the bare system Hamiltonian $H_S$, as expected. Fig. \ref{fig:1}(a) displays $\gamma$ as well as the amount of renormalization
\begin{equation}
\Delta := \varepsilon^* -\varepsilon
\label{eq:delta}
\end{equation}
which is independent of the bare spacing $\varepsilon$,  as functions of the coupling strength $\chi$, for $|\chi| < \Omega$. These quantities are displayed in non-dimensional units where energies are rescaled by $\varepsilon$. As $|\chi|/\varepsilon$ approaches the stability limit $\pm \Omega/\varepsilon$, $\Delta/\varepsilon$ and $\gamma/\varepsilon$ diverge, while they vanish as the coupling $\chi/\varepsilon$ approaches zero.
From Eq. (\ref{eq:varepsilon*}), we note that, given certain values of the spacing $\varepsilon$ and of $\Omega$, there exists a value of $\chi$ for which the renormalized energy spacing $\varepsilon^*$ vanishes, meaning that an effective degeneracy of the qubit is induced by the presence of the oscillator.
In Fig. \ref{fig:1}(b), $\Delta/\varepsilon$ and $\gamma/\varepsilon$ are plotted as functions of $\Omega/\varepsilon$, for fixed $\beta$ and $\chi$, and for $\Omega/\varepsilon > |\chi|/\varepsilon$.

The graphs in Fig.  \ref{fig:1} correspond to  values of $\Omega$ and $\chi$ that match the regime of values used in an experimental implementation  of the model Hamiltonian in Eq. (\ref{eq:H-time-ind}) with superconducting circuits, as  it has been recently reported \cite{Schuster:2007ph}.
In that experiment $|\chi|$ is about two orders of magnitudes smaller than $\Omega$, thus the leading corrections to the energy spacing are of first order and those of the shift $\gamma$ are at most of second order in $\chi/\varepsilon$.
\begin{figure}
\begin{center}
\includegraphics[width=8.5cm]{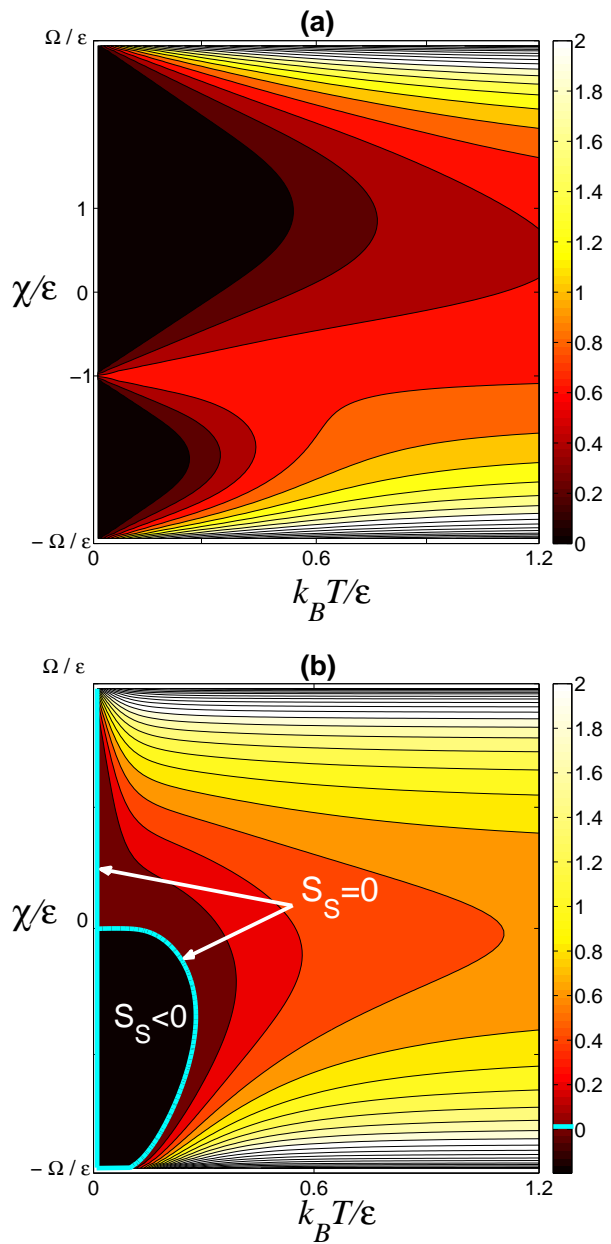}
 \caption{Contour plot of dimensionless entropy $S_S/k_B$, Eq. (\ref{eq:Ss}), as a function of dimensionless temperature $k_B T/\varepsilon= 1/(\beta \varepsilon)$, and dimensionless interaction strength $\chi/\varepsilon$, for (a) $\Omega/\varepsilon =3 $ (top panel), (b) $\Omega/\varepsilon =1/3 $ (bottom panel). The entropy is nowehere negative in panel (a). In panel (b),  it assumes negative values in the region, labelled as $S_S<0$, enclosed by the level line $S_S=0$ (thick light blue line).}
\label{fig:2}
\end{center}
\end{figure}

At low temperatures we find the following limiting results:
\begin{eqnarray}
\lim_{\beta \rightarrow \infty}\gamma &=&0 \label{eq:LimGamma} \\
\lim_{\beta \rightarrow \infty}\varepsilon^* &=& \varepsilon + \chi \label{eq:LimVarepsilon*}\\
\lim_{\beta \rightarrow \infty}\frac{\partial^k}{\partial \beta^k}\varepsilon^* &=& 0 \qquad k=1,2,3 ... \label{eq:LimDersVarepsilon*}
\end{eqnarray}
From the previous two equations we deduce that the degeneracy of the spectrum occurs at $T=0$ for the special value $\chi= - \varepsilon$. In the following we will come back to the effect of this degeneracy on the system's entropy and specific heat.

\subsection{Thermodynamic Entropy}

From the partition function, $Z_S$, one obtains the free energy:
\begin{equation}
F_S = -k_B T \ln Z_S = -(1/\beta) \ln Z_S,
\end{equation}
and the entropy:
\begin{equation}
S_S = -\frac{\partial F_S}{\partial T} = k_B \beta^2 \frac{\partial  F_S}{\partial \beta} \ .
\label{eq:Ss}
\end{equation}
In Fig. \ref{fig:2}(a) the entropy following from Eq. (\ref{eq:Z_S(t)}) with $\varepsilon(t)=\varepsilon$, is displayed as a function of dimensionless temperature $k_B T/\varepsilon= 1/(\beta \varepsilon)$ and dimensionless coupling strength, $\chi / \varepsilon$ for a fixed value of rescaled oscillator energy quantum $\Omega/\varepsilon$, larger than $1$. As $\chi/\varepsilon$ approaches the instability values $\pm\Omega/\varepsilon$ the entropy diverges. For all values of  $\chi/\varepsilon$, the entropy vanishes at zero temperature in agreement with the third law. An exception is at the special case $\chi/\varepsilon=-1$, where the ground state assumes a finite degeneracy. Put differently, for $\chi/\varepsilon=-1$, the zero temperature entropy is no longer zero but assumes the finite positive value $k_B \ln 2$ \cite{Planck1910,Giauque1928,Giauque1933,Pauling1935}. This $k_B \ln 2$ term is a consequence of the fact that in the limit of zero temperature and for $\chi=-\varepsilon$, the effective spacing $\varepsilon^*$ and all its higher order derivatives with respect to temperature vanish (see Eq. (\ref{eq:LimVarepsilon*},\ref{eq:LimDersVarepsilon*})).
At finite temperatures there are  values of $\chi$ for which the spacing  $\varepsilon^*$ vanishes, however these do not coincide with the values of $\chi$ for which the entropy is $k_B \ln 2$, since then the first derivative of $\varepsilon^*$ with respect to $\beta$ does not vanish and consequently yields a contribution to the entropy.

\begin{figure}
\begin{center}
\includegraphics[width=8.5cm]{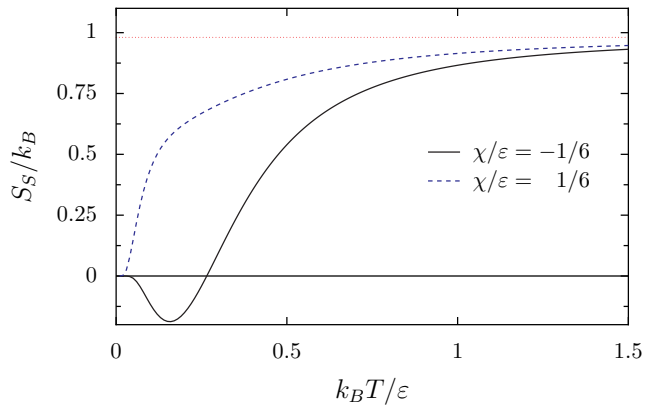}
 \caption{Dimensionless entropy $S_S/k_B$ as a function of dimensionless temperature $k_B T/\varepsilon= 1/(\beta \varepsilon)$, for $\Omega/\varepsilon=1/3$, and  $\chi/\varepsilon=-1/6 $ (solid line), $\chi/\varepsilon=1/6 $ (dashed line). The dotted line is the asymptotic value calculated from Eq. (\ref{eq:HighTempEntr}).}
\label{fig:3}
\end{center}
\end{figure}

Fig. \ref{fig:2}(b) depicts the entropy as a function of dimensionless temperature $k_B T/\varepsilon= 1/(\beta \varepsilon)$ and dimensionless coupling strength, $\chi / \varepsilon$, for a fixed value of rescaled oscillator energy quantum $\Omega/\varepsilon$, smaller than $1$. The most prominent feature in this case $\Omega/\varepsilon<1$ is the appearance of a region of \emph{negative entropy} for small values of $k_B T$ and negative coupling (the region labeled as $S_S<0$ in Fig. \ref{fig:2}(b)). Interestingly, the experiment reported in Ref. \cite{Schuster:2007ph} lies in this region $\Omega<\varepsilon$, $\chi <0$ where the entropy may become negative. For the parameter values reported therein, a negative entropy is expected below $\sim 22$ mK.
From Fig. \ref{fig:2}(b) we notice that, for positive $\chi$, the entropy vanishes at absolute zero temperature, in accordance to the third law of thermodynamics \cite{Planck1910}, and reaches a plateau at high temperatures, without becoming negative. For negative $\chi$, it vanishes as well at zero temperature, however with increasing low temperatures, entropy first decreases until it reaches a negative minimum, and  then increases until it approaches a positive plateau-value at high temperatures.
This behaviour is further illustrated in Fig. \ref{fig:3}.

Independently of the sign of $\chi$, at high temperature the entropy  reaches the asymptotic value:
\begin{equation}
\lim_{\beta\rightarrow 0} S_S = k_B \ln
\left(\frac{2\Omega^2}{\Omega^2-\chi^2} \right)
\label{eq:HighTempEntr}
\end{equation}
which notably does not depend on the spacing $\varepsilon$. For $\chi=0$,
the high temperature entropy in Eq. (\ref{eq:HighTempEntr})
becomes equal to $k_B \ln 2$, reflecting the fact that spin up and
spin down states become equally populated at infinite temperature.
For $\chi \neq 0$, it assumes values larger than $k_B \ln 2$, and diverges for $|\chi|$ approaching $\Omega$.

\begin{figure}
\begin{center}
\includegraphics[width=8.5cm]{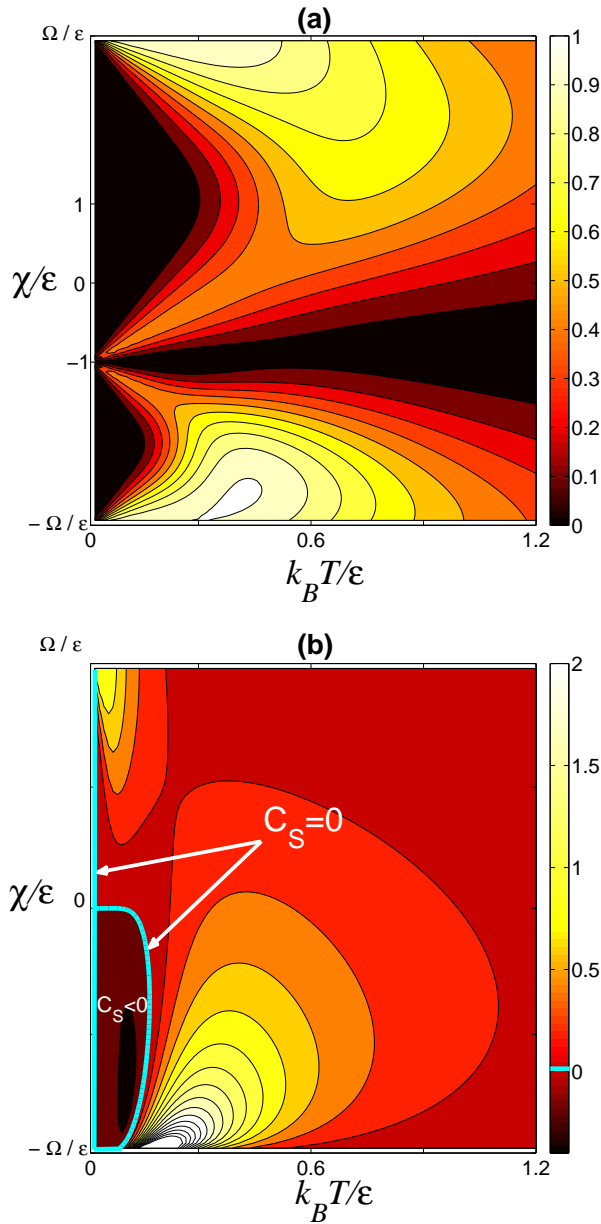}
 \caption{Contour plot of dimensionless specific heat $C_S/k_B$, Eq. (\ref{eq:Cs}), as a function of dimensionless temperature $k_B T/\varepsilon= 1/(\beta \varepsilon)$, and dimensionless interaction strength $\chi/\varepsilon$, for (a) $\Omega/\varepsilon =3 $ (top panel), (b) $\Omega/\varepsilon =1/3 $ (bottom panel). The specific heat is nowehere negative in panel (a). In panel (b), it assumes negative values in the region, labelled as $C_S<0$, enclosed by the level line $C_S=0$ (thick light blue line).}
\label{fig:4}
\end{center}
\end{figure}


\subsection{Specific heat}

From the entropy (\ref{eq:Ss}) one obtains the the specific heat of the open two level system:
\begin{equation}
C_S = T \frac{\partial S_S}{\partial T}=-\beta \frac{\partial S_S}{\partial \beta}
\label{eq:Cs}
\end{equation}
Figs. \ref{fig:4}(a) and\ref{fig:4}(b) represent the specific heat as a function of dimensionless temperature $k_B T/\varepsilon= 1/(\beta \varepsilon)$ and dimensionless coupling strength, $\chi / \varepsilon$ for fixed values of rescaled oscillator energy quantum $\Omega/\varepsilon$.

\begin{figure}[t]
\begin{center}
\includegraphics[width=8.5cm]{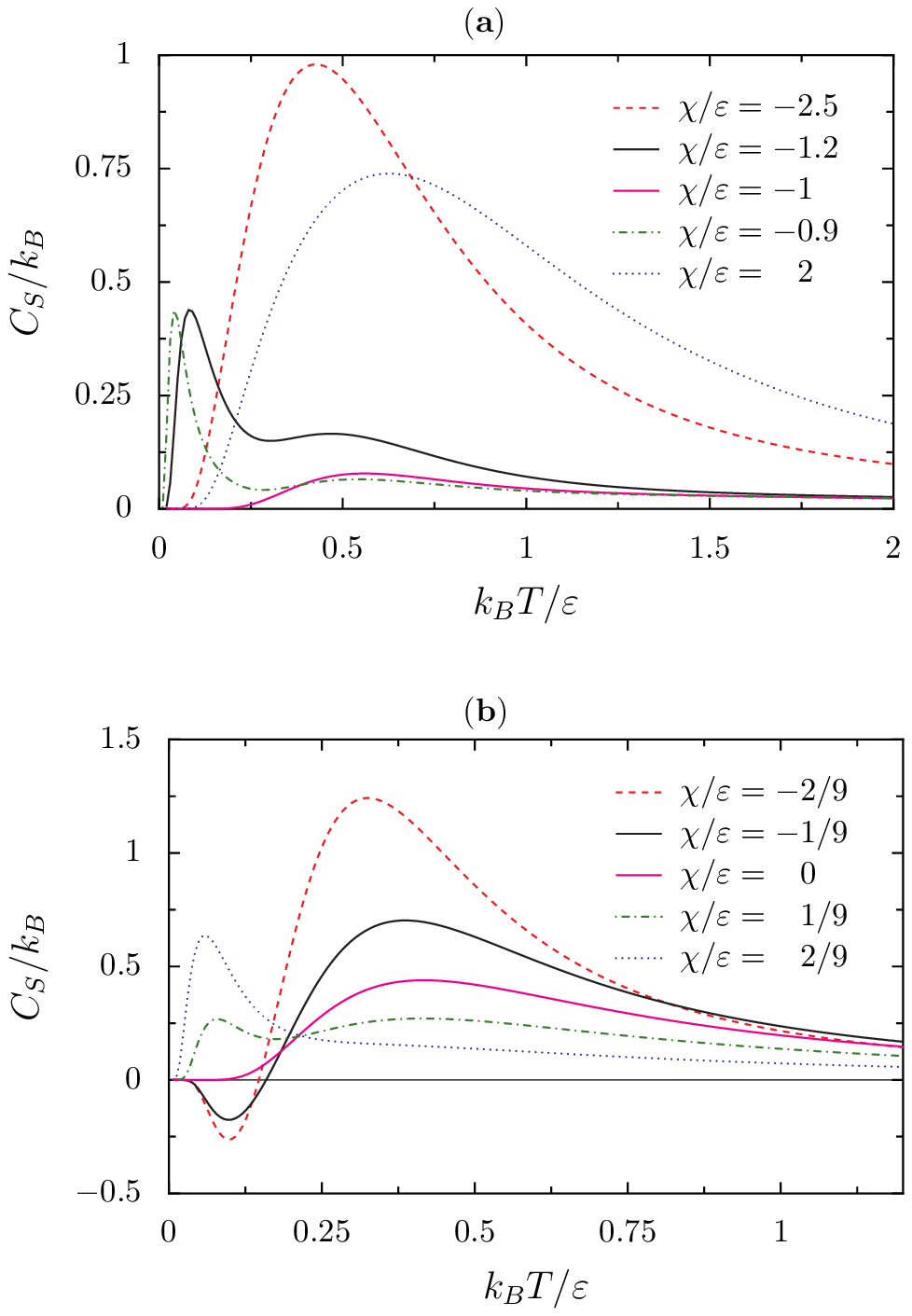}
 \caption{Dimensionless specific heat $C_S/k_B$ as a function of rescaled temperature $k_B T/\varepsilon=1/(\beta \varepsilon)$, for various values of $\chi/ \varepsilon$ and (a) $\Omega/\varepsilon=3$ (top panel), (b) $\Omega/\varepsilon=1/3$ (bottom panel).}
\label{fig:5}
\end{center}
\end{figure}

In Fig. \ref{fig:4}(a) $\Omega$ is larger than $\varepsilon$.
For values  $| \chi/ \varepsilon|<\Omega/\varepsilon$, the specific heat vanishes at zero temperature. With growing temperatures, it first increases, then reaches a maximum and finally goes again to zero. The maximum occurs at decreasing temperatures as $\chi/ \varepsilon \rightarrow-1$. As  $\chi/ \varepsilon$ approaches $-1$  we also see the appearance of a second maximum for larger values of $k_B T$. These features are further illustrated in Fig. \ref{fig:5}(a).

The specific heat landscape changes drastically for
$\Omega<\varepsilon$, Fig. \ref{fig:4}(b). The most
relevant feature in this parameter range is the appearance of a
region of \emph{negative specific heat} at low temperatures and
negative $\chi/\varepsilon$ (the region labelled as $C_S<0$ in Fig.
\ref{fig:4}(b)). From Fig. \ref{fig:4}(b)
we observe that, for $0< \chi/ \varepsilon < \Omega/\varepsilon$ the
specific heat starts from zero at zero temperature, reaches a
maximum and goes to zero again at high temperatures. For
$-\Omega/\varepsilon< \chi / \varepsilon < 0$ the specific heat
starts at $T=0$ from zero, reaches a negative minimum with
increasing temperature, then a positive maximum and finally goes to
zero at high temperature. These features are further illustrated in
Fig.  \ref{fig:5}(b). From Fig. \ref{fig:5}(b) we also
notice that the curves corresponding to positive $\chi$ all cross
each other within a very small temperature range around $k_B
T/\varepsilon \sim 0.21$  for the given value $\Omega/\varepsilon =
1/3$. Indeed, for  $k_B T/\varepsilon = 0.21$,  $\Omega/\varepsilon
= 1/3$, and $\chi/ \varepsilon$ ranging from $0$ to
$\Omega/\varepsilon$, the specific heat is almost constant (with
variations within 5 \% of its value). An analogous situation happens
also for other values of  $\Omega/\varepsilon <1$, showing that in
this regime one should expect the presence of a narrow temperature
range for which the specific heat is not very sensitive to changes
in coupling strength $\chi$, as long as this remains positive.
\begin{figure}[t]
\begin{center}
\includegraphics[width=8.5cm]{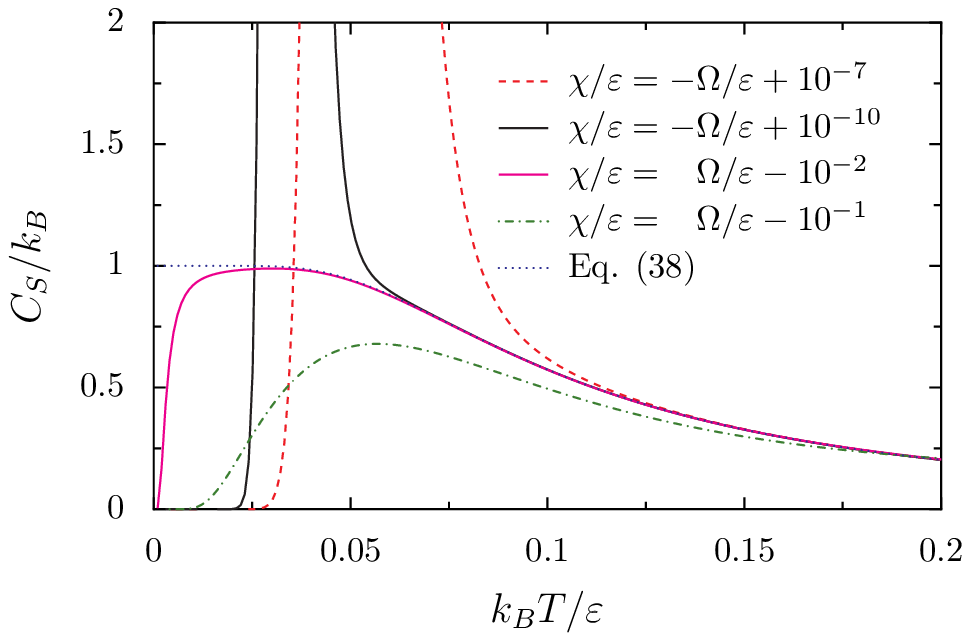}
 \caption{Dimensionless specific heat $C_S/k_B$ as a function of rescaled temperature $k_B T/\varepsilon=1/(\beta \varepsilon)$ for $\Omega/\varepsilon =1/3$ and various small values of  $|\chi/\varepsilon| - \Omega/\varepsilon$.}
\label{fig:6}
\end{center}
\end{figure}
Regardless of whether $\Omega$ is larger or smaller than $\varepsilon$, the specific heat approaches a unique functional form in the limits
as $\chi$ approaches $\pm \Omega$. This limiting function can be calculated analytically:
\begin{equation}
\lim_{\chi \rightarrow \pm \Omega} C_S = k_B \left(1- \left[
\frac{2k_B T}{\Omega}\sinh \left( \frac{\Omega}{2k_B T}
\right)\right]^{-2}\right) \label{eq:limCsChi2Omega}
\end{equation}
The fact that it does not tend to zero at zero temperature is not in contrast with the third law of thermodynamics, since for $\chi =\pm \Omega$ the system is no longer stable.

Fig. \ref{fig:6} depicts the behavior of the specific heat as $\chi/\varepsilon \rightarrow \pm \Omega/\varepsilon$ for $\Omega/\varepsilon <1$. The convergence to the limiting function in Eq. (\ref{eq:limCsChi2Omega}) as $\chi/\varepsilon \rightarrow +\Omega/\varepsilon$ is quite fast  compared to the much slower convergence in the other limit $\chi/\varepsilon \rightarrow -\Omega/\varepsilon$.  The approach to the limit is qualitatively very different in the two cases. In both cases the specific heat vanishes at low temperature and it approaches the limiting curve in Eq. (\ref{eq:limCsChi2Omega}) for large temperatures. However in the case $\chi/\varepsilon \sim -\Omega/\varepsilon$, the specific heat displays a drastic peak at intermediate temperatures. As $\chi/\varepsilon$ approaches $-\Omega/\varepsilon$,  the peak becomes increasingly pronounced while getting closer to the origin of the temperature axis. In the limit $\chi/\varepsilon \rightarrow -\Omega/\varepsilon$, eventually the peak becomes a delta singularity at zero temperature. This singularity contributes with a finite term $\lambda$ to the total heat $Q=\int C_S dT$, which, in analogy with first order phase transitions, can be interpreted as a latent heat.

\section{\label{sec:conclusions}Conclusions}
We illustrated the validity of the Jarzynski equality and the work fluctuation theorem in the strong coupling regime, for the model Hamiltonian (\ref{eq:H}). The central role is played by the thermodynamic partition function of the open system, that incorporates the interaction of the system of interest with its environment. The influence of the interaction is of major importance even in the seemingly trivial case in which the system bath  interaction Hamiltonian commutes with both the bath and the system Hamiltonians, notwithstanding claims to the contrary \cite{Teifel:2007qr}. We computed the Hamiltonian of mean force for this model explicitly and studied its equilibrium thermodynamics. In particular we discussed its entropy and its specific heat as functions of temperature and other system parameters. Like for other strongly coupled system \cite{HIT_NJP08,Ingold:2008mz} these two quantities can become negative at low temperature.
Despite entropy and specific heat may become negative, they vanish at zero temperature, in accordance with the third law of thermodynamics. The only exception to this, is for the special value of coupling strength $\chi$ exactly equal to $-\varepsilon$, for which the zero temperature entropy is equal to $k_B \ln 2$. This result is, however, not in contradiction  with the third law but rather corroborates this law; this is so because the two level fluctuator becomes degenerate in this case, as is clearly indicated by the Hamiltonian of mean force.

Interestingly, recent experiments in circuit cavity quantum electrodynamics \cite{Schuster:2007ph} use a parameter regime  where negative entropy and specific heat may appear. For the architecture presented in \cite{Schuster:2007ph}, these are expected below $\sim 22$ mK and $\sim 20$ mK respectively. In cavity quantum electrodynamics the Hamiltonian in Eq. (\ref{eq:H-time-ind}) is obtained from the time-independent Jaynes-Cummings  model Hamiltonian in the rotating wave approximation \emph{and} dispersive regime \cite{Schleich}. These conditions imply weak coupling $|\chi| \ll \Omega,\varepsilon$, which  in fact is the case for Ref.  \cite{Schuster:2007ph}. Whether a Hamiltonian of the type in Eq. (\ref{eq:H}), with time- dependent $\varepsilon(t)$, and/or possibly strong coupling  can be implemented with superconducting circuits remains an open problem.

\section*{Acknowledgements} The authors wish to thank David Zueco for fruitful discussions. Financial support by the German
Excellence Initiative via the {\it Nanosystems Initiative Munich}
(NIM) and the Volkswagen Foundation (project I/80424) is gratefully
acknowledged.


\end{document}